\documentclass[prl,twocolumn,showpacs]{revtex4}
\usepackage{amsmath}
\usepackage[dvips]{graphicx}
\usepackage{amssymb}

\begin{document}
\title{Sodium Bose-Einstein Condensates in an Optical Lattice}
\author{K. Xu\footnote{Contact Info: kwxu@mit.edu}, Y. Liu, J.R. Abo-Shaeer, T. Mukaiyama, J.K. Chin, D.E. Miller and W. Ketterle\footnote{Website: cua.mit.edu/ketterle\_group}}
\affiliation{Department of Physics, MIT-Harvard Center for
Ultracold Atoms, and Research Laboratory of Electronics, MIT,
Cambridge, MA 02139}
\author{Kevin M. Jones}
\affiliation{Physics Department, Williams College, 33 Lab Campus Dr.,
Williamstown, MA 01267}
\author{Eite Tiesinga}
\affiliation{Atomic Physics Division, National
Institute of Standards and Technology, 100 Bureau Drive, Stop 8424,
Gaithersburg, Maryland 20899, USA}
\date{\today}

\begin{abstract}
The phase transition from a superfluid to a Mott insulator  has been observed in a $^{23}$Na Bose-Einstein condensate. A dye laser detuned $\approx 5$~nm red of the Na $3^2$S$ \rightarrow 3^2$P$_{1/2}$ transition was used to form the three dimensional optical lattice. The heating effects of the small detuning as well as the three-body decay processes constrained the timescale of the experiment. Certain lattice detunings were found to induce a large loss of atoms.  These loss features were shown to be due to photoassociation of atoms to vibrational levels in the Na$_2$ $(1) ^3\Sigma_g^+$ state.
\end{abstract}


\maketitle

\section*{Introduction}
Optical lattices have become a powerful tool to enhance the effects of interaction in ultracold atomic systems to create strong correlations and probe many-body physics beyond the mean-field theory \cite{catalotti01, orzel01, greiner02, greiner02colrev, paredes04, schori04, kohl05}. Simply through varying the depth of the lattice potential, one changes the tunneling rate as well as the on-site interaction energy by changing the confinement of the atoms. The strength of the atomic interaction can be directly tuned with a magnetic Feshbach resonance \cite{inou98}. In comparison to $^{87}$Rb, which has been used in almost all experiments on optical lattices, $^{23}$Na has stronger and wider Feshbach resonances that are experimentally accessible \cite{sten98stro, marte02}. One such resonance has been used to produce quantum degenerate Na$_2$ molecules \cite{xu03}. Therefore, a sodium condensate loaded into an optical lattice would be a rich and flexible system for studying strong correlations.

So far, most optical lattice experiments have been performed with relatively heavy atomic species (e.g. rubidium and potassium) for which the recoil frequencies are lower and lasers are readily available to achieve trap depths of several tens of recoil frequencies at a few tens of milliwatts. For $^{23}$Na, high power single-mode lasers are necessary for similar experiments. In this work, we chose to use a dye laser red-detuned by $\approx 5$ nanometers from the D lines of sodium (589~nm). The spontaneous scattering rate limited the time window of the experiment to less than 50~ms, but was still sufficient to satisfy the adiabaticity condition to explore the quantum phase transition from a superfluid to a Mott insulator. We also observed strong atom losses at various lattice laser detunings, which were interpreted as photoassociation transitions. The particular molecular states responsible for these transitions were identified through theoretical calculations and previous experimental data.

\section*{Experiment Setup}
A $^{23}$Na Bose-Einstein condensate containing up to $10^6$ atoms in the $\left | F=1, m_F = -1 \right \rangle$ state was first produced in a magnetic trap and subsequently loaded into a crossed optical dipole trap. The optical trap was derived from a single-mode 1064~nm infrared laser, with the horizontal and vertical beams detuned by 60~MHz through acousto-optic modulators. The number of condensed atoms was varied through three-body decay in a tight trap ($\omega_{x,y,z} = 2 \pi \times 200, 328, 260$~Hz) , after which the trap was decompressed ($\omega_{x,y,z} = 2 \pi \times 110, 155, 110$~Hz) to allow further evaporation and re-thermalization. A vertical magnetic field gradient was applied to compensate for gravity and avoid sagging in the weaker trap.

A dye laser operated at 594.710~nm was used to set up a three dimensional optical lattice. The three beams were focused to $1/e^2$-waist of $\sim 82$~$\mu$m at the condensate, and retro-reflected to form standing waves. The two horizontal beams were orthogonal to each other, while the third beam was slanted at $\sim 20^{\circ}$ with respect to the vertical axis due to limited optical access. The three beams were frequency-shifted by $\pm 30$~MHz and 80~MHz to eliminate cross interference between different beams.

The gaussian profile of the lattice beams added an additional harmonic trapping potential, while the localization of atoms at the lattice sites increased the repulsive mean field interaction. At the maximum lattice depth, the trap frequencies due to the combined potentials of the optical dipole trap and the lattice beams were $\sim 510$~Hz for all three dimensions. The trap parameters were chosen such that during the ramping of the optical lattice potential, the overall size of the cloud (parametrized by Thomas-Fermi radii) remained approximately constant in order to minimize intra-band excitations (the mean Thomas-Fermi radius is $\sim 14~\mu$m for $10^6$ atoms). The peak per-lattice-site occupancy numbers achieved in our experiment were between 3 to 5.

\section*{Quantum Phase Transition}
Atoms held in a shallow optical lattice can tunnel freely from site to site and form a superfluid phase. As the lattice is made deeper,  the atomic interaction is increased while the tunneling rate between lattice sites is exponentially suppressed. The system then undergoes a phase transition to an insulating phase -- the Mott-insulator -- in which each lattice site contains a definite fixed number of atoms. According to the mean-field theory for the homogenous systems of atoms in the lowest band of an optical lattice, the critical point for the phase transition from a superfluid to a Mott-insulator state with $n$ atoms per lattice site is determined by \cite{fisher89, krauth92, jaksch98}:
\begin{equation}
U =z( 2n+1+2\sqrt{n(n+1)} ) J
\label{eq1}
\end{equation}
where:
\begin{equation}
U  =  \frac{4 \pi \hbar^2 a_s}{m} \int d^3x |w(x)|^4
\label{eq2}
\end{equation} 
is the on-site interaction energy;
\begin{equation}
J =  \int d^3x \, w^{*}(x-\lambda_{latt}/2)(-\frac{\hbar^2}{2 m} \nabla^2+V_{latt}(x))w(x)
\label{eq3} 
\end{equation}
is the tunneling rate between adjacent lattice sites; $z$ is the number of nearest neighbors in the lattice (6 for a cubic lattice); $m$ is the atomic mass; $a_s$ is the $s$-wave scattering length (2.75~nm for $^{23}$Na)); $w(x)$ is the Wannier function; $\lambda_{latt}$ is the lattice wavelength; $V_{latt}(x)$ is the lattice potential.

\begin{figure}[htbp]
\includegraphics[width=80mm]{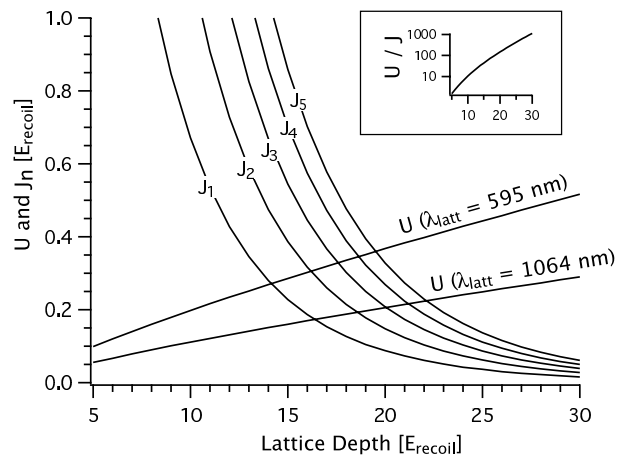}
\caption{Determination of the phase transition points: The phase transition at occupancy number $n$ occurs when $J_n = z (2n+1+2\sqrt{n(n+1)}) J$ as given by equation (\ref{eq1}) equals $U$. The horizontal location of the crossing point where $J_n=U$ gives the critical lattice depth. The inset shows the ratio of $U$ over $J$ as a function of the lattice depth.}
\label{uandj}
\end{figure}

Figure~\ref{uandj} shows $U$ and $J_n = z (2n+1+2\sqrt{n(n+1)}) J$ for a cubic lattice as a function of the lattice depth, obtained through a band-structure calculation. All energies are expressed in units of the recoil energy $E_{recoil}=\hbar^2 k_{latt}^2/2m$, where $k_{latt}=2\pi/\lambda_{latt}$ is the lattice wavenumber. With this scaling $J$ is independent of $\lambda_{latt}$. The peak occupancy number in our experiment was $\lesssim 5$. From Fig.~\ref{uandj}, we find that the the critical points are at a lattice depth of 14.2, 16.2, 17.6, 18.7, and 19.5 (all in units of $E_{recoil}$) for $n=1, 2, 3, 4,$ and 5 respectively. The inset of Fig.~\ref{uandj} shows that the ratio of $U/J$ increases rapidly with increasing lattice depth.

When a weak harmonic trap is present in addition to the lattice potential, as is the case for the experiment, the atomic density is not uniform. Nevertheless, Eqs. (\ref{eq1} -- \ref{eq3}) can be used to estimate the lattice depth needed to observe the Mott-insulator phase transition at any point in the harmonic trap. Given the local density of the initial condensate, a local value of $n$ can be estimated and thus the local critical lattice depth can be read off from Fig.~\ref{uandj}. Since the critical depth increases with $n$, one expects that as the lattice depth is increased, shells of different occupancies will undergo the transition to the Mott-insulator phase starting from the edge of the density profile and moving in towards the center.

\begin{figure*}[htbp]
\includegraphics[width=6in]{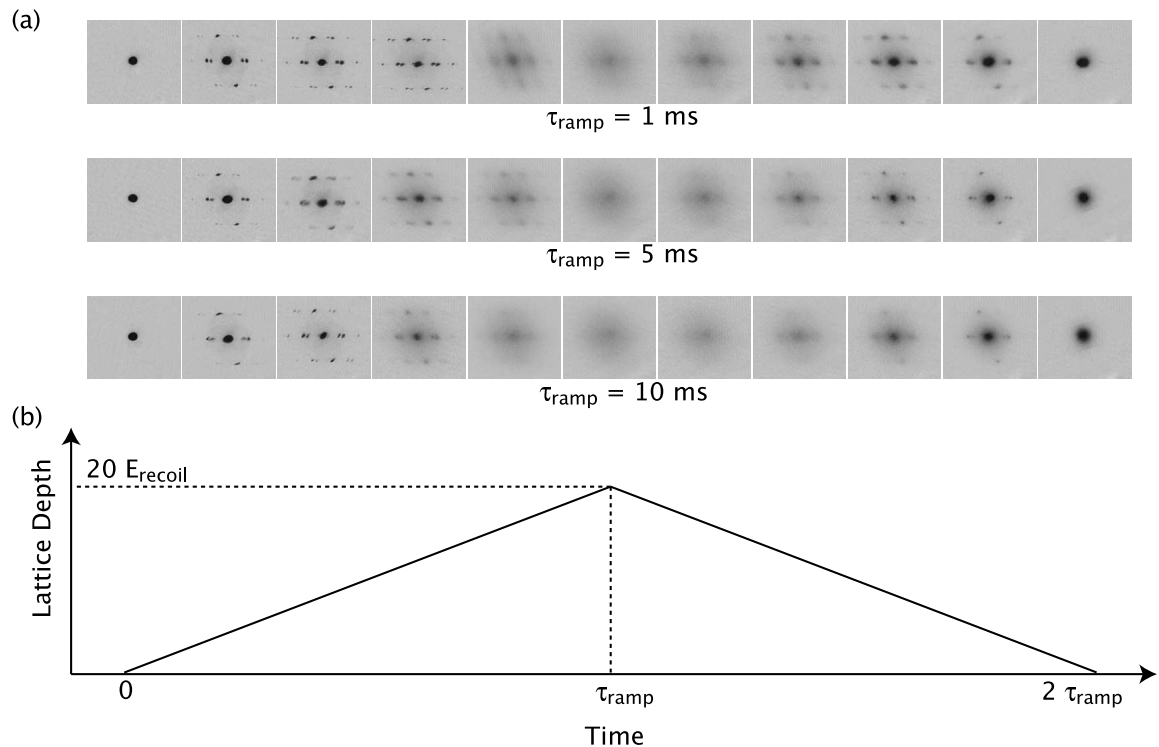}
\caption{ (a) Observation of the superfluid to Mott insulator transition: The lattice depths for the sequence of images are from left to right 0, 4, 8, 12, 16, 20, 16, 12, 8, 4, 0~$E_{recoil}$. The time-of-flight is 7~ms and the field of view is $1000~\mu$m$\times 1200~\mu$m. The peak occupancy number $n$ is 5 for these images, and the phase transition should occur between 14 and 20~$E_{recoil}$ according to the mean-field calculation. (b) Time dependence of the lattice depth.}
\label{supmott}
\end{figure*}

In our experiment, the optical lattice was linearly ramped up to a maximum potential of $20~E_{recoil}$ in a variable time $\tau_{ramp}$ ($E_{recoil}= h \, 24.4$~kHz for our system). The lattice depth was calibrated by probing the energy difference between the first and the third band at zero quasi-momentum with small amplitude modulation of the lattice beams (see, e.g., \cite{schori04}). After reaching the peak value, the lattice was ramped back down again in $\tau_{ramp}$. The ramp sequence was stopped at different times when both the trap and the lattice were suddenly switched off (in $\lesssim 1~\mu$s). Absorption images were then taken after some time-of-flight as shown in Figure \ref{supmott}. The disappearance of the interference pattern as the lattice depth was increased indicated the loss of phase coherence and a transition from the superfluid state to the Mott insulator state \cite{greiner02}. The subsequent revival of the interference patterns as the lattice depth was reduced ensured that the system remained essentially in the ground state during the ramping process. Different $\tau_{ramp}$'s were used to check the adiabaticity condition. The peak spontaneous light scattering rate was about 21~s$^{-1}$ at the maximum intensity. Therefore for $\tau_{ramp} \leq 10$~ms, less than 20\% of the atoms spontaneously scattered a photon.

After the lattice was fully ramped down, most of the atoms ($> 80\%$) remained in the condensed fraction while the rest were heated and distributed across the first Brillouin zone. Based on the number of atoms that remained in the condensate after the lattice was fully ramped down,  we conclude that $\tau_{ramp}\gtrsim 1$~ms satisfies the intra-band adiabaticity condition. In the following discussion, all measurements were performed for $\tau_{ramp} = 1,5,10$~ms, but only the data for $\tau_{ramp} = 5$~ms are shown as representative of similar results unless otherwise noted.

\begin{figure}[htbp]
\includegraphics[width=80mm]{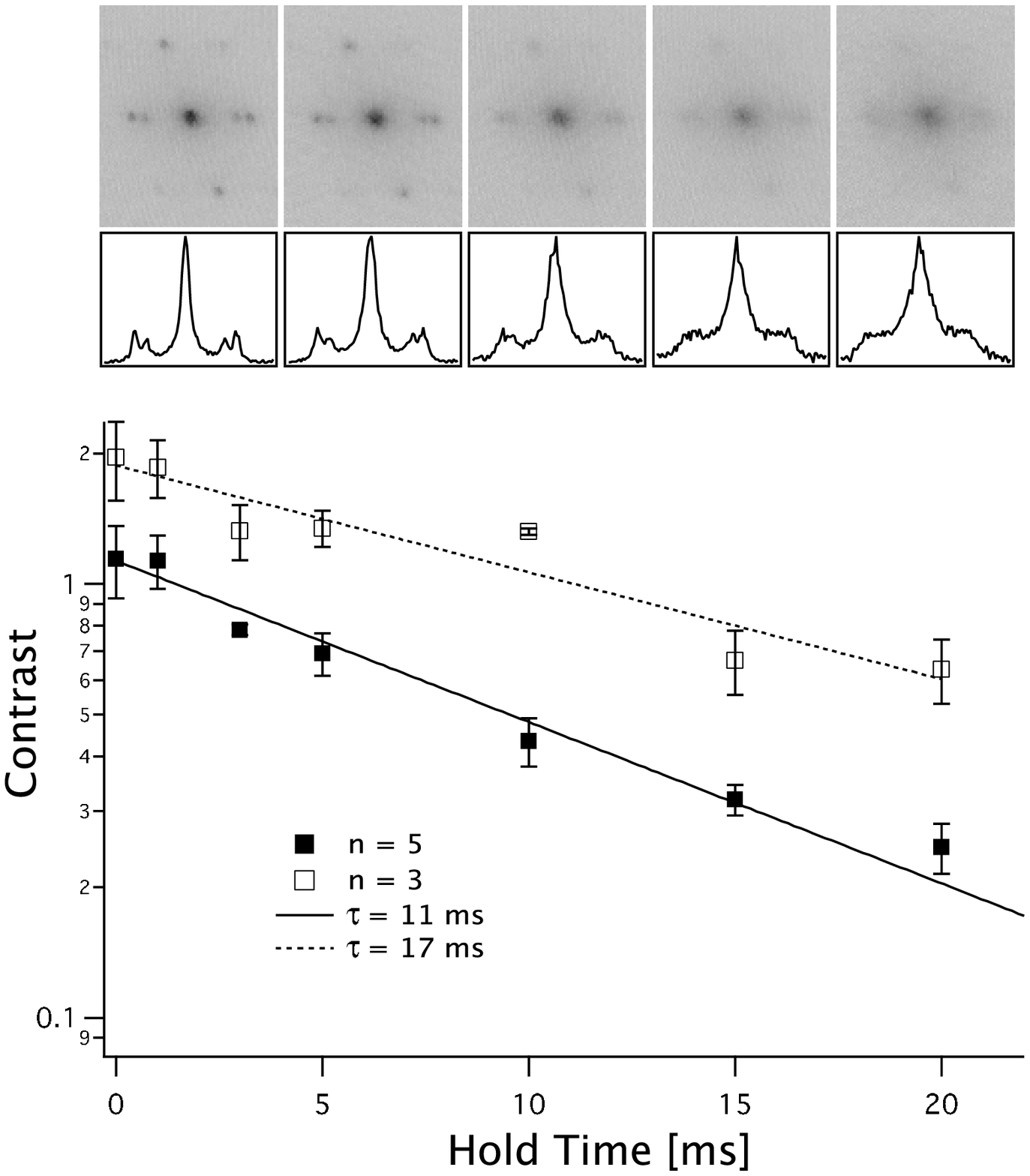}
\caption{Lifetime of the Mott insulator state: The top row of absorption images were taken for hold times 0, 5, 10, 15, 20~ms. The second row is the horizontal cross section of the atomic density profile taken at the center of the absorption images. The decay of the Mott insulator state is observed as the loss of contrast of the interference peaks. The curves are exponential fit. The error bars in the graph are statistical standard deviation of multiple shots.}
\label{lifetime}
\end{figure}

To characterize the lifetime of the Mott-insulator phase, we held the lattice depth at the maximum level for various amounts of time before ramping the lattice down to $8~E_{recoil}$ (below the Mott-insulator transition point) and taking the time-of-flight image. If the system remains in the ground state, the contrast of the interference pattern should be recovered, whereas additional heating populates the Brillouin zone and reduces the interference contrast. A cross-section of the density profile was taken along the horizontal direction showing the interference peaks on top of a broad background (see Figure \ref{lifetime}). The 5 interference peaks and the broad background were fit by 6 gaussians. The ratio between the total integrated area of the peaks and the background was used as the contrast to quantify the heating of the system. The contrast gives a more sensitive measure of the heating compared to simply counting the recovered condensate atoms. As the atoms in the interference peaks quickly move apart, they are not as broadened by the mean-field expansion as a single condensate.

We performed the same measurement for two different peak occupancy numbers $n \simeq 3$ and 5. Figure \ref{lifetime} shows the decay of the contrast and the lifetime $\tau$ was determined using an exponential fit. The fitting error on the lifetime was less than 17\%. The lifetime was about 50\% longer for $n=3$, implying that inelastic collision processes significantly contributed to the heating of the system. The three-body decay rate at the maximum lattice potential for the peak on-site atomic density ($\sim 10^{16}$~cm$^{-3}$ for $n=5$) is about 100~s$^{-1}$ \cite{stamp98odt}, consistent with the observed lifetimes of $\sim 10$~ms. The peak density of a condensate in a harmonic trap and therefore the peak occupancy number scales with $2/5$ power of the total number of atoms, and our method for varying the number of atoms (through three body decay) was unable to consistently produce low enough atom numbers for peak occupancy $\leq 2$. The signal-to-noise ratio of our current imaging system also became marginal for such low atom numbers.

 \section*{Photoassociation Resonances}
\begin{figure}[htbp]
\includegraphics[width=80mm]{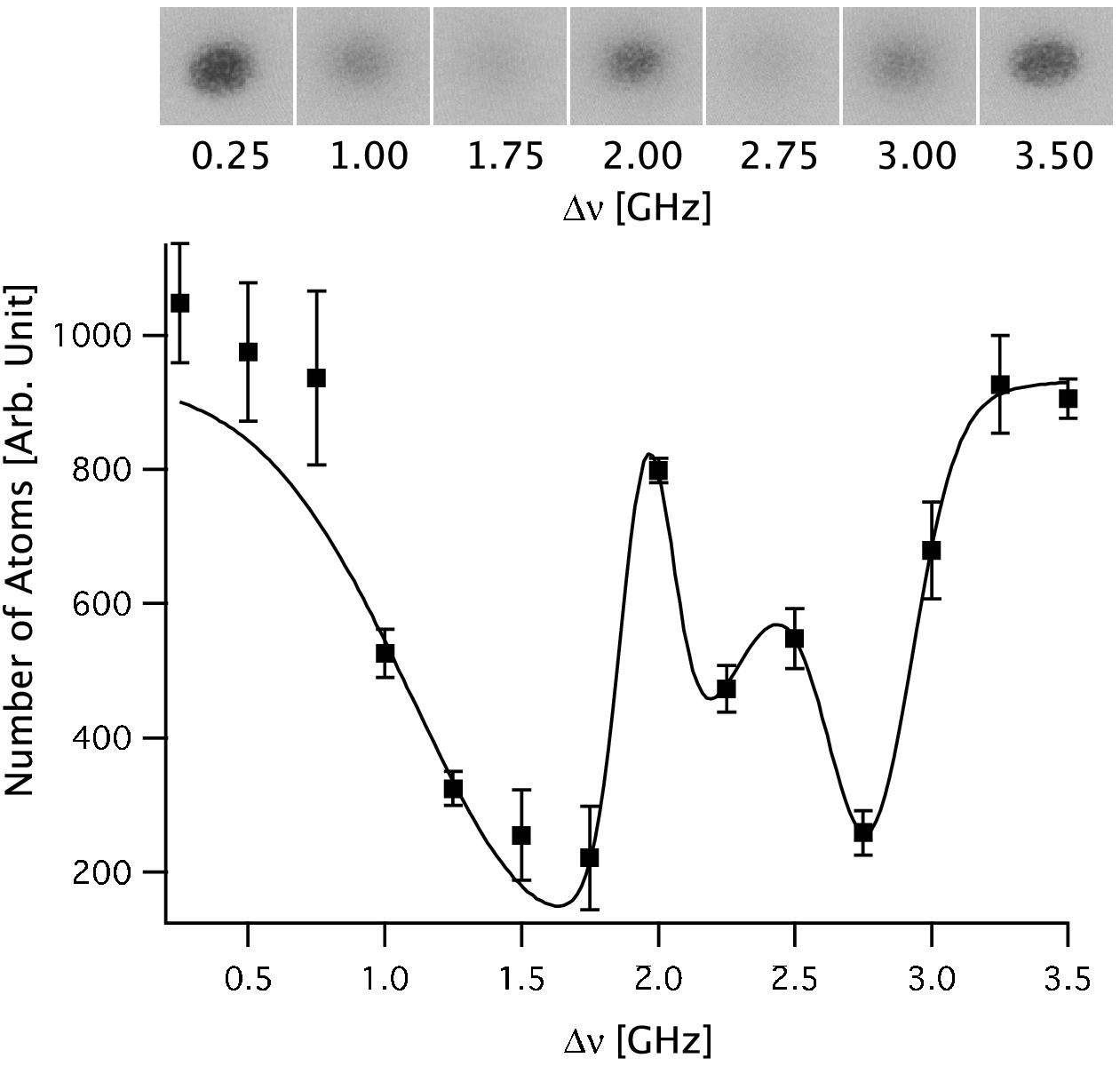}
\caption{Photoassociation resonances: Atom loss from the optical lattice as a function of the lattice laser detuning.  The top row of images shows the remaining atom cloud after the lattice was turned off and the cloud expanded for 30~ms. The lower graph shows the remaining number of atoms for these and additional frequency points.  The zero of the $x$ axis corresponds to a laser wavelength of 594.490 nm as determined with a wavemeter with accuracy $\sim1$ GHz.  The error bars are one standard deviation uncertainty.  The line is only a guide and does not have a theoretical underpinning.}
\label{photoassoc}
\end{figure}

In this experiment, in addition to losses due to three body recombination, we observed large losses of atoms for certain specific tunings of the lattice laser in the range 592~nm to 595~nm. A sample of such a loss feature is shown in Fig.~\ref{photoassoc}. For this measurement, the same ramp sequence was used as before with $\tau_{ramp}=1$~ms.  The peak intensity is about 280~W/cm$^2$ in each lattice beam. Due to the intentional frequency shifts between the three lattice beams the effective bandwidth of the lattice light as seen by the atoms is $\approx 100$~MHz.  For the narrow frequency scan range of Fig.~\ref{photoassoc}, the relative frequency scale was determined to better than 25~MHz using a Fabry-Perot cavity with a 2~GHz free spectral range.

Single-photon photoassociations proved to have caused these losses. The lattice laser is tuned by $\approx160$ cm$^{-1}$ to the red of the atomic $3^2$S$ \rightarrow 3^2$P$_{3/2}$ transition and thus is in a spectral region where it might drive photoassociation transitions \cite{Weiner1999, Jones2006, Stwalley1999} to rovibrational levels in molecular states dissociating to either the $3^2$S$ +3^2$P$_{1/2}$ or $3^2$S$ +3^2$P$_{3/2}$ limits. Such a photoassociation transition,  followed by the spontaneous radiative decay of the excited molecule into either a bound ground electronic-state molecule or into ``hot'' atoms, results in significant losses of atoms from the lattice. It is therefore important to identify the locations and strengths of these resonances and choose the appropriate lattice wavelength to avoid such losses.

\begin{figure}[htbp]
\includegraphics[width=80mm]{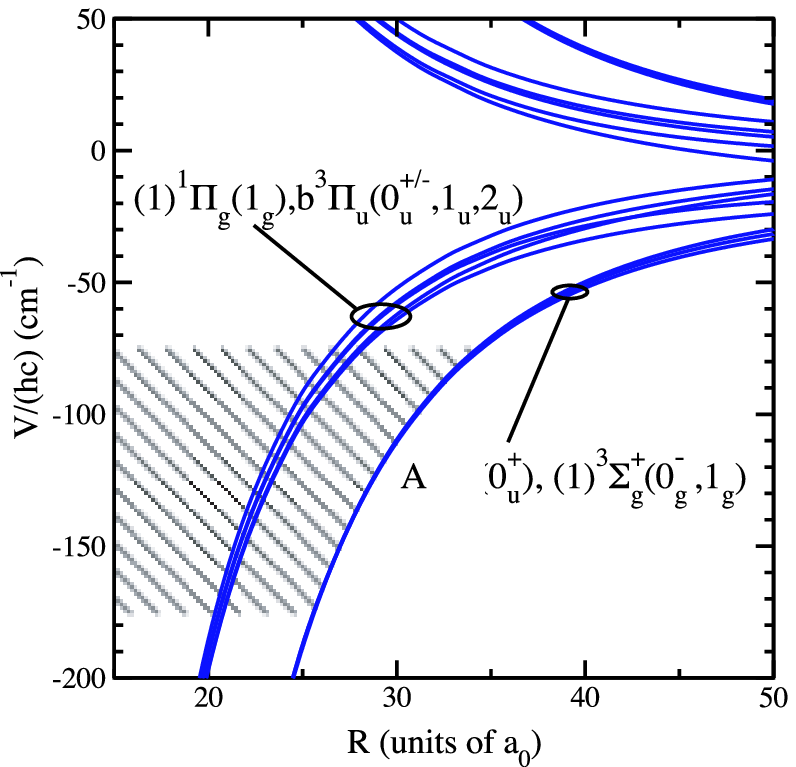}
\caption{Long range molecular potentials near the Na$_2$ $3^2$S$+3^2$P separated atom limit (1 $a_0$ = 0.0529~nm):  The zero of the vertical scale is the $3^2$S$+3^2$P$_{3/2}$ limit, the $3^2$S$+3^2$P$_{1/2}$ limit lies 17.2~cm$^{-1}$ lower. The hatched region indicates the energy range explored in the experiment.  The potentials are labeled by both Hund's case (a) state labels of the form $^{2S+1}\Lambda^\pm_{g/u}$ and by Hund's case (c) labels $\Omega^\pm_{g/u}$ given in parenthesis after the case (a) label.  For small $R$ the curves labeled by a given case (a) label become degenerate, while for larger $R$,  as the effect of spin-orbit coupling becomes increasingly important, the degeneracy is lifted. }
\label{pots}
\end{figure}

There is an extensive body of knowledge on the photoassociation of ultracold alkali-metal atoms and the behavior of the molecular potentials dissociating to the $3^2$S$ +3^2$P$_{1/2}$ or $3^2$S$ +3^2$P$_{3/2}$ limits \cite{Weiner1999, Jones2006, Stwalley1999}. Figure \ref{pots} shows the relevant excited molecular potentials as a function of internuclear separation $R$. The ground electronic states of Na$_2$ are the X$^1\Sigma_g^+$ and a$^3\Sigma_u^+$ states and two colliding ground state atoms will be some mixture of these symmetries.  To the extent that the excited states are well described as $\Sigma$ or $\Pi$ states, the $g \leftrightarrow u$ and $\Delta S=0$ selection rules imply that photoassociation transitions are allowed only to the two $\Sigma$ states (A$^1\Sigma_u^+$ and (1)$^3\Sigma_g^+$).

\begin{figure*}[htbp]
\includegraphics[width=160mm]{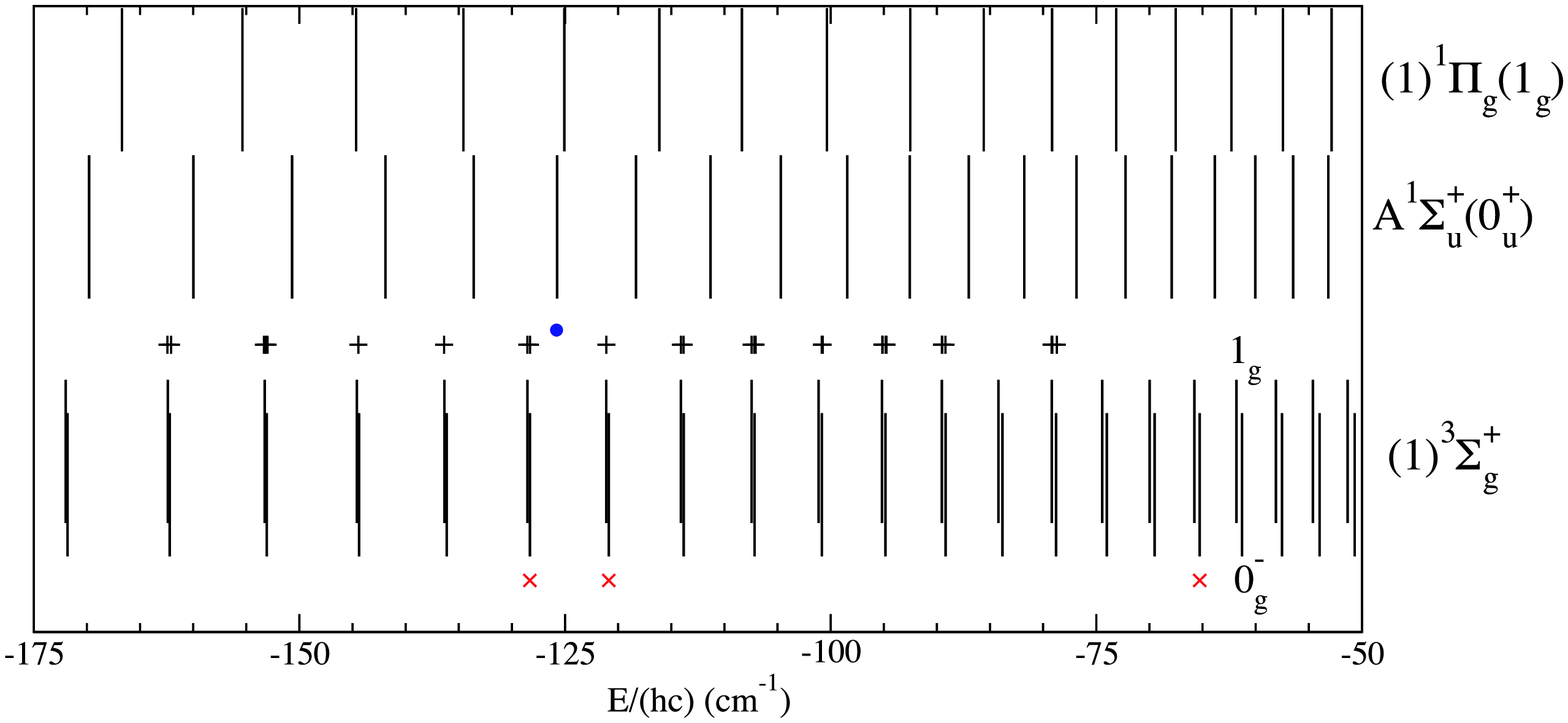}
\caption{Comparison of loss features (plusses and dot) observed in a Bose condensate trapped in an optical lattice, to known molecular level positions:   The $x$ axis is the binding energy of the molecular level relative to the $3^2$S$(f=2)+ 3^2$P$_{3/2}(f=3)$ atomic asymptote.  If a loss feature, observed at a laser photon energy of $E_{\rm photon}$, is due to photoassociation of two $3^2$S$(f=1)$ Na atoms to a bound molecular level,  then the molecular level is bound by $E_0- E_{\rm photon}$ where $E_0/(hc)$=16973.4636 cm$^{-1}$.  The laser tunings of observed loss features are plotted in this way (the pluses and the dot). The lines labeled $(1)^1\Pi_g (1_g)$ are binding energies of the $J=1$ rotational levels of vibrational levels in this potential measured in a magnetooptical trap (MOT) (\cite{ratliff94} and unpublished work).  The lines labeled A$^1\Sigma_u^+ (0_u^+)$ are the energies reported in the RKR curve of \cite{Tiemann1996} (nominally $J=0$ positions) shifted to agree with the photoassociation measurement of \cite{mckenzie02} corrected for rotation. The lines labeled (1)$^3\Sigma_g^+$ are from an extended Movre-Pichler model calculation calibrated to experimental data (the crosses) from a two-color photoassociation experiment in a MOT.  The lines displaced upwards (downwards) are the $1_g$ ($0_g^-$) component of the (1)$^3\Sigma_g^+$ state. The crosses indicate the three calibration points which lie in this spectral range and are measurements of the $J=0$ levels of the $0_g^-$ component.}
\label{compare}
\end{figure*}

Previous experiments have identified the locations of the strong transitions to the A$^1\Sigma_u^+$ state \cite{Tiemann1996} and the weak transitions to the $(1)^1\Pi_g (1_g)$ state \cite{ratliff94}.  These are shown in Fig.~\ref{compare}. We looked for but failed to find any significant losses attributable to the weak $(1)^1\Pi_g (1_g)$ state resonances. We were able to confirm one of the A-state resonances, indicated by the dot in Fig.~\ref{compare}. Since the primary goal of the present work was to avoid photoassociation losses we did not investigate known A-state locations further. Our search focused on the strong resonances due to the (1)$^3\Sigma_g^+$ state which had not been previously observed in this spectral region. In 0.5~GHz steps, we scanned through a 30~GHz range around the theoretically predicted locations based on the model and auxiliary experimental data discussed below.  In all but one such scans, we were able to observe between 1 to 3 dips in the remaining atom numbers within a range of $\lesssim 15$ GHz, including the loss feature shown in Fig.~\ref{photoassoc}. As shown in Fig.~\ref{compare} the agreement of the observed locations with the predictions and auxiliary measurements confirms that these loss features are due to photoassociation to the (1)$^3\Sigma_g^+$ state. The locations of the rovibrational levels of the b$^3\Pi_u$ state are not known in the current tuning range. Their spacing, however, should be equal to that of the $(1)^1\Pi_g$ levels and our observations are not consistent with such spacings.

The potential curves used in the calculation of the (1)$^3\Sigma_g^+$ vibrational levels were generated from an extended version of the model developed by Movre and Pichler for calculating the combined effects of the $1/R^3$ resonant dipole interaction and the atomic spin-orbit interaction.  Such models have been extensively used to interpret photoassociation experiments \cite{Weiner1999, Jones2006, Stwalley1999}. To the long range potentials generated by the Movre-Pichler type calculation we append the results of {\it ab initio} calculations on the short range molecular potentials (in the chemical bonding region). These short range potentials are not sufficiently accurate to allow predictions of absolute vibrational positions.  It is necessary to make slight adjustments to these potentials to match experimentally measured vibrational positions, which were obtained through a separate photoassociation spectroscopy experiment in a dark-spot magnetooptical trap (MOT) containing Na $3^2$S$(f=1)$ atoms at about 300 $\mu$K, using a two color ionization scheme and an ion detector \cite{jones97, Amelink2000}. The photoassociation spectra taken in the MOT have higher resolution than the loss features in the lattice experiment.

In the spectral region of interest for the lattice experiment, three vibrational levels of the (1)$^3\Sigma_g^+$ were identified by measurements in the MOT.   The spectra show a $1_g$ component with a complicated hyperfine/rotational pattern, and, slightly higher ($\sim 0.3$ cm$^{-1}$) in energy, a $0_g^-$ component with a simpler nearly rotational pattern.  This ordering of the $1_g$ and $0_g$ components is in agreement with the Movre-Pichler model.  For these levels, the $J=0$ feature of the $0_g^-$ component were found to be at 16845.155 cm$^{-1}$, 16852.585 cm$^{-1}$, and 168908.201 cm$^{-1}$, calibrated to iodine lines with an estimated uncertainty of about 0.004 cm$^{-1}$ and are shown in Fig.~\ref{compare}. This new data on the (1)$^3\Sigma_g^+$  was sufficient to calibrate our extended Movre-Pichler model without any further adjustable parameters. The predicted positions shown in Fig.~\ref{compare} agree well with those of the loss features observed in the lattice experiment, thus identifying the loss features as  photoassociation transitions to levels in the  (1)$^3\Sigma_g^+$ state.

Based on this insight, we chose a wavelength of  594.710~nm for our lattice experiment (corresponding to -158.5 cm$^{-1}$ in Fig.~\ref{compare}).  This tuning lies $>45$~GHz from the closest molecular resonance.  Given the observed on-resonance photoassociation rate of 1~ms$^{-1}$ in the lattice, photoassociative decay can be ignored at such detunings as the rate scales as the square of the ratio of the natural linewidth to the detuning.

\section*{Conclusions}
In this paper, we explored the possibility of using a dye laser detuned $\approx 5$~nm from the Na D lines to study many-body physics of a sodium BEC in an optical lattice, which could allow for an independent control of interaction using magnetic Feshbach resonances. The superfluid to Mott-insulator transition was observed in a Na Bose-Einstein condensate for the first time. The main technical difficulties are due to the heating from the spontaneous light scattering and three-body decay processes. In addition, several photoassociation resonances were observed and identified by means of auxiliary spectroscopy measurements combined with theoretical modeling. These resonances were avoided by choosing an appropriate lattice wavelength. In future experiments, we plan to use a high-power single-mode infrared laser at 1064~nm to eliminate atomic light scattering (Figure \ref{uandj} also shows the transition points for a 1064~nm lattice). Moreover, heating from three body recombination can be avoided by using occupancy numbers less than three.

\section*{Acknowledgement}
The authors would like to thank Fredrik Fatemi and Paul Lett for their help in obtaining the MOT photoassociation data used here, and Widagdo Setiawan for experimental assistance. We also thank Carl Williams for initiating the communications regarding the photoassociation resonances. This research is supported by NSF, ONR, ARO, and NASA.

\end{document}